\documentclass{article}

\usepackage{PRIMEarxiv}

\usepackage[utf8]{inputenc} % allow utf-8 input
\usepackage[T1]{fontenc}    % use 8-bit T1 fonts
\usepackage{hyperref}       % hyperlinks
\usepackage{url}            % simple URL typesetting
\usepackage{booktabs}       % professional-quality tables
\usepackage{amsfonts}       % blackboard math symbols
\usepackage{nicefrac}       % compact symbols for 1/2, etc.
\usepackage{microtype}      % microtypography
\usepackage{lipsum}
\usepackage{fancyhdr}       % header
\usepackage{graphicx}       % graphics
\usepackage{subcaption}
\graphicspath{{media/}}     % organize your images and other figures under media/ folder

\title{Segmentation-free integration of nuclei morphology and spatial transcriptomics for retinal images
}

\author{
  Eduard Chelebian\thanks{eduard.chelebian@it.uu.se} , Carolina Wählby  \\
  Dept. of Information Technology and SciLifeLab \\
  Uppsala Univeristy \\
  Uppsala, Sweden
   \And
  Pratiti Dasgupta, Zain Samadi, Amjad Askary \\
  Dept. of Molecular, Cell and Developmental Biology \\
  University of California, Los Angeles \\
  Los Angeles, CA, USA
}

\begin{document}
\maketitle

\begin{abstract}

This study introduces SEFI (SEgmentation-Free Integration), a novel method for integrating morphological features of cell nuclei with spatial transcriptomics data. Cell segmentation poses a significant challenge in the analysis of spatial transcriptomics data, as tissue-specific structural complexities and densely packed cells in certain regions make it difficult to develop a universal approach. SEFI addresses this by utilizing self-supervised learning to extract morphological features from fluorescent nuclear staining images, enhancing the clustering of gene expression data without requiring segmentation. We demonstrate SEFI on spatially resolved gene expression profiles of the developing retina, acquired using multiplexed single molecule Fluorescence In Situ Hybridization (smFISH). SEFI is publicly available at 
\url{https://github.com/eduardchelebian/sefi}.
\end{abstract}

% keywords can be removed

\section{Introduction}

Imaging-based spatial transcriptomics techniques, such as seqFISH \cite{eng2019transcriptome}, MERFISH \cite{chen2015spatially}, and \textit{in situ} sequencing \cite{ke2013situ}, enable high-resolution detection of spatial gene expression. Analyzing these data typically involves segmenting nuclei in DAPI-stained tissue to enable single-cell measurements. However, cell segmentation is challenging due to factors such as the three-dimensional structure of cells, tissue-specific differences, and the presence of densely packed areas, which may require additional staining for accurate identification \cite{thomas2017review}.

In contrast, segmentation-free approaches leverage the intrinsic distribution of detections for aggregation \cite{park2021cell,andersson2024points2regions}. These aggregated detections are then clustered to identify cell types or cell niches. While DAPI imaging serves as a basis for segmentation, it also contains valuable morphological information that remains underutilized. Unlike sequencing-based spatial transcriptomics, which have developed integration methods with H\&E staining \cite{hu2021spagcn}, there are currently no established methods for integrating imaging-based spatial transcriptomics with DAPI.

To effectively integrate imaging-based spatial transcriptomics measurements with DAPI staining results, it is essential to extract morphological features from the images. This can be achieved through self-supervised learning, which we have previously demonstrated can define meaningful regions \cite{chelebian2024self}. However, existing integration methods do not fully leverage these representations \cite{chelebian2024makes}.

In this study, we introduce SEFI (SEgmentation-Free Integration) for nuclei morphology and imaging-based transcriptomics. Utilizing self-supervised learning, we extract morphological representations from DAPI images. This approach enhances the segmentation-free clustering of genes by incorporating relevant morphological components, thereby improving the accuracy of niche detection. 

We apply our method to images of developing retina, where segmentation is particularly challenging due to the densely populated neuroblastic layers. Cells in this context exhibit distinct morphological features according to their cell state and developmental stage. By employing SEFI, we aim to refine the classification of retinal cell types beyond what gene expression alone can provide. 

\section{Method}
\begin{figure}[htbp]
    \centering
    \includegraphics[width=\linewidth]{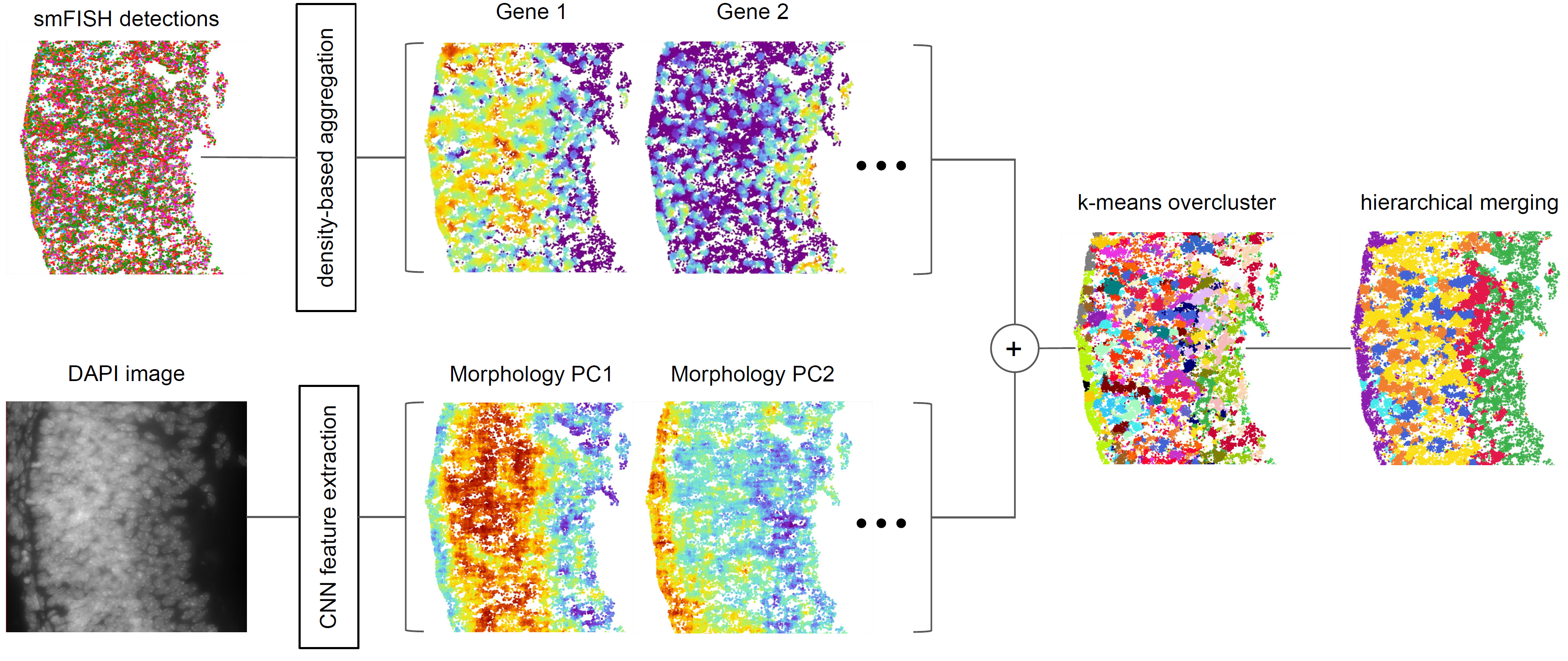}
    \caption{Steps of SEFI, from left to right, starting with creating density maps for each gene, and morphological maps by learned feature extraction. The dimensions of the morphological maps are reduced by PCA. Finally, niches are detected by k-means pixel clustering and hierarchical cluster merging.}
    \label{fig:workflow}
\end{figure}

SEFI consists of three main steps, as illustrated in Figure \ref{fig:workflow}. First, we generate individual gene expression maps from spatial transcriptomics detections. Next, we perform feature extraction on DAPI nuclei morphology images using convolutional neural networks (CNN). Finally, we cluster the joint gene and morphological features using k-means, followed by the merging of clusters using hierarchical clustering.

\textbf{Gene expression maps}. To generate individual gene expression maps from spatial transcriptomics detections, we employ the density-based aggregation proposed in Points2Regions by Andersson \textit{et al.} \cite{andersson2024points2regions}. This method calculates local composition features from point cloud data, focusing on regions with similar detection classes. 

\textbf{Morphology maps}. Similar to the approach in \cite{chelebian2024self}, we use a pretrained ResNet18 model with SimCLR \cite{chen2020simple} self-supervision to extract features from DAPI nuclei morphology images. The resulting 512-dimensional features are reduced using PCA, retaining 95\% of the variance.

\textbf{Clustering and merging}. The gene expression maps and reduced morphological maps are used as inputs for clustering. We first perform k-means clustering based on the number of genes, followed by hierarchical clustering to progressively merge clusters based on a defined stopping criterion.

\section{Experiments and results}

\subsection{Data description}
The dataset consists of three multiplexed smFISH experiments conducted on developing E14 mouse retinas, using a panel of 33 genes selected based on their variable expression among retinal progenitor cells. The original DAPI images are shown in Appendix \ref{app:images}.

\subsection{Morphological features improve clustering}
\begin{figure}[htbp]
    \centering
    \includegraphics[width=0.5\linewidth]{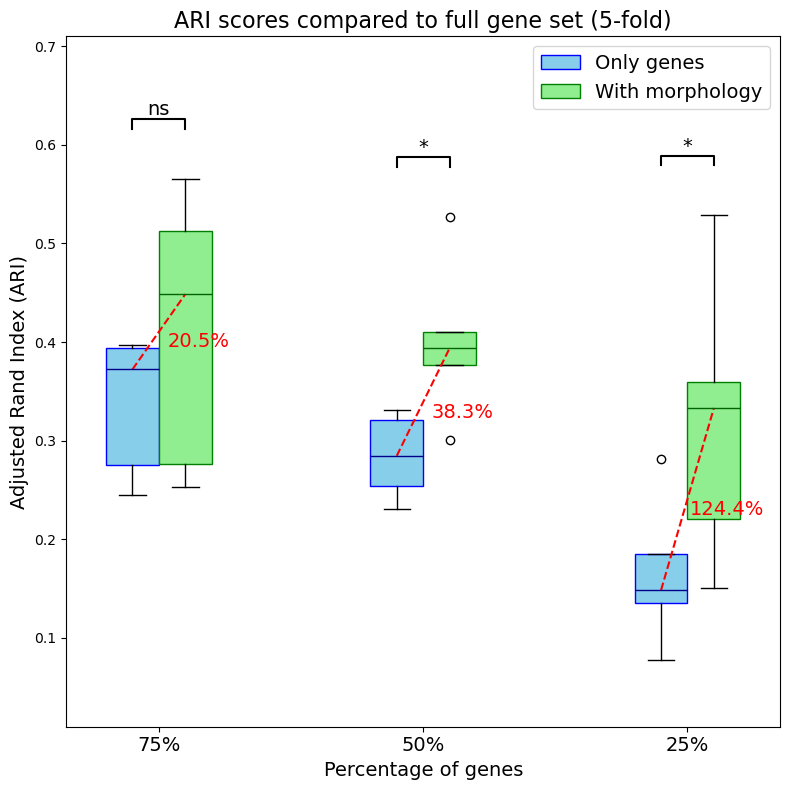}
    \caption{Morphological information can compensate for information lost when including fewer genetic markers for niche detection.}
    \label{fig:comparison}
\end{figure}

Since defining a ground truth for these experiments is challenging, we devised a framework to assess the usefulness of morphological features. Our hypothesis was that clustering using all 33 genes without additional data would provide the most representative result. We used this clustering outcome as a proxy for the ground truth. Then, we progressively removed genes at random and compared the Adjusted Rand Index (ARI) \cite{hubert1985comparing} from clustering with just the remaining genes and with the added morphological features. As shown in Figure \ref{fig:comparison}, reducing the number of genes makes the addition of morphological features increasingly important, helping the ARI score approach the clustering result from using all 33 genes. This suggests that, in targeted methods, where not all relevant genes may be included, morphological data could help compensate for missing genetic information.

\subsection{Biological relevance of the clusters}
\begin{figure}[htbp]
    \centering
    \includegraphics[width=.9\linewidth]{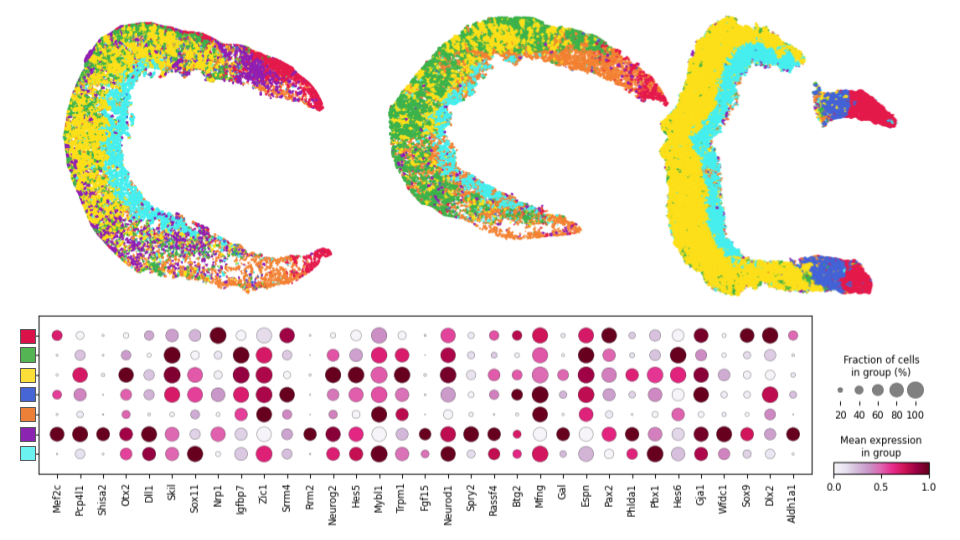}
    \caption{Results from applying the full SEFI pipeline on the 33 genes and the morphology}
    \label{fig:results}
\end{figure}

To ensure that the clusters identified by SEFI are biologically relevant, we applied the full pipeline using both the 33 genes and morphological features and calculated the average gene expression within each cluster. Figure \ref{fig:results} demonstrates the robustness of SEFI, showing that clusters consistently correspond to distinct regions of the retina. For example, cluster 7 (cyan) is enriched with ganglion cell markers such as SOX11 \cite{jiang2013transcription} and PBX1 \cite{zagozewski2014role} across all samples. Ciliary margin cells are predominantly found in cluster 1 (red), marked by GJA1 \cite{calera2006connexin43}, although other markers like WFDC1 and ALDH1A1 are inconsistently distributed across multiple clusters. Finally, clusters 2 (green) and 3 (yellow) are more heterogeneous, containing a mix of markers from photoreceptor, and bipolar cells \cite{kaufman2019transcriptional}.  

\section*{Acknowledgments}
This work was supported by an ERC Consolidator grant (CoG 682810) to C.W., a Liljewalch travel scholarship to E.C., and a National Eye Institute grant to A.A. (R00EY031782).

%Bibliography
\bibliographystyle{unsrt}  
\bibliography{references}  

\clearpage

\appendix

\section*{Appendix}
\renewcommand{\thesubsection}{\Alph{subsection}}

\subsection{Original images}
\label{app:images}

\begin{figure}[htbp]
\centering
\begin{subfigure}{0.32\textwidth}
    \includegraphics[width=\textwidth]{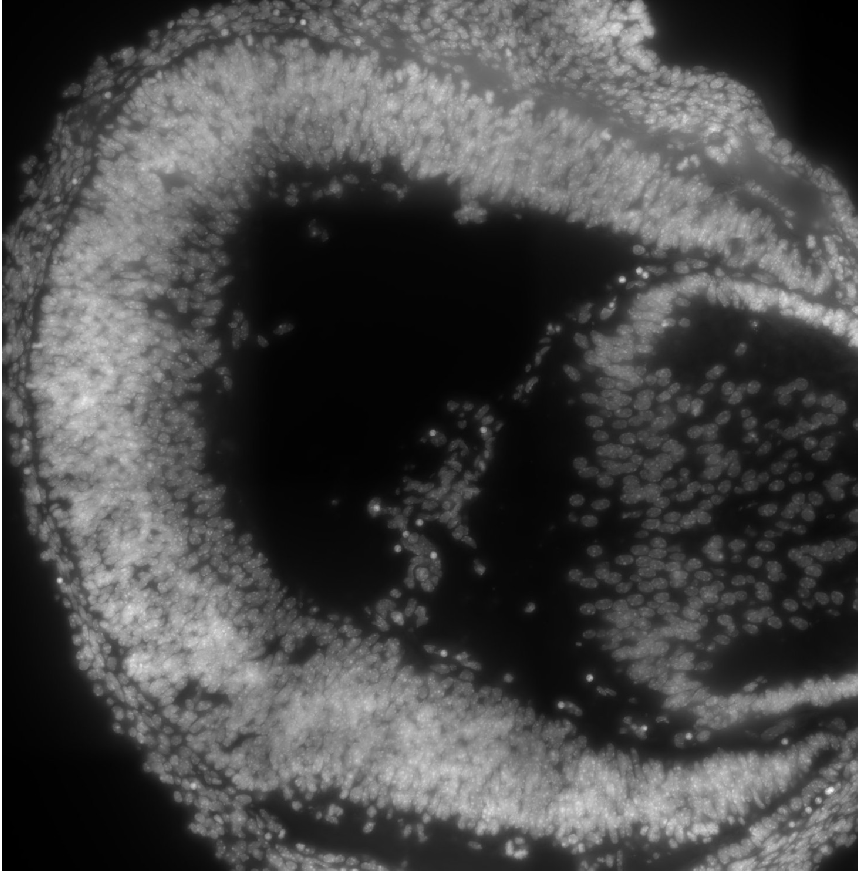}
    \caption{Eyeball sample 1}
    \label{fig:first}
\end{subfigure}
\hfill
\begin{subfigure}{0.32\textwidth}
    \includegraphics[width=\textwidth]{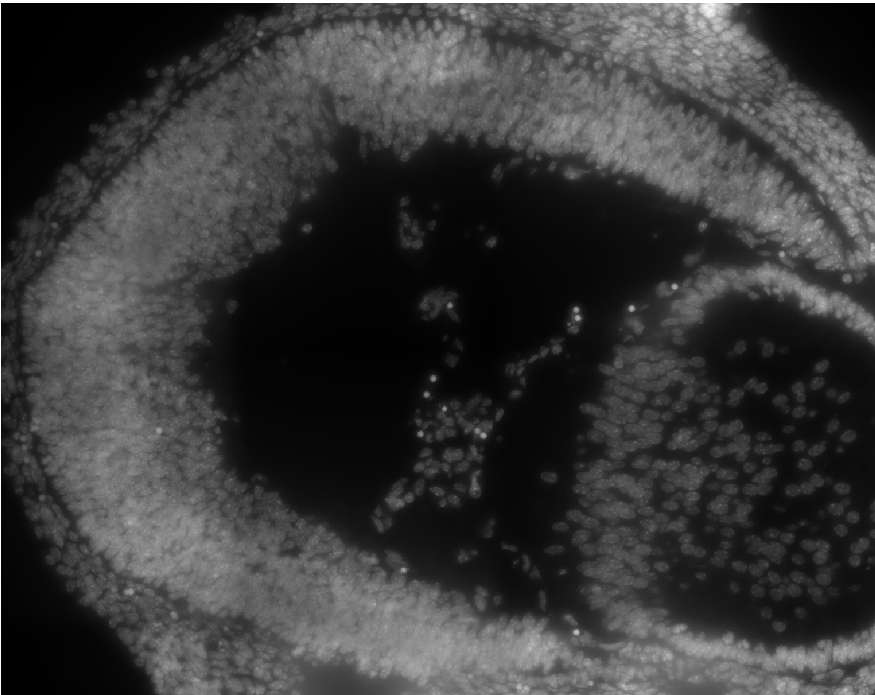}
    \caption{Eyeball sample 2}
    \label{fig:second}
\end{subfigure}
\hfill
\begin{subfigure}{0.32\textwidth}
    \includegraphics[width=\textwidth]{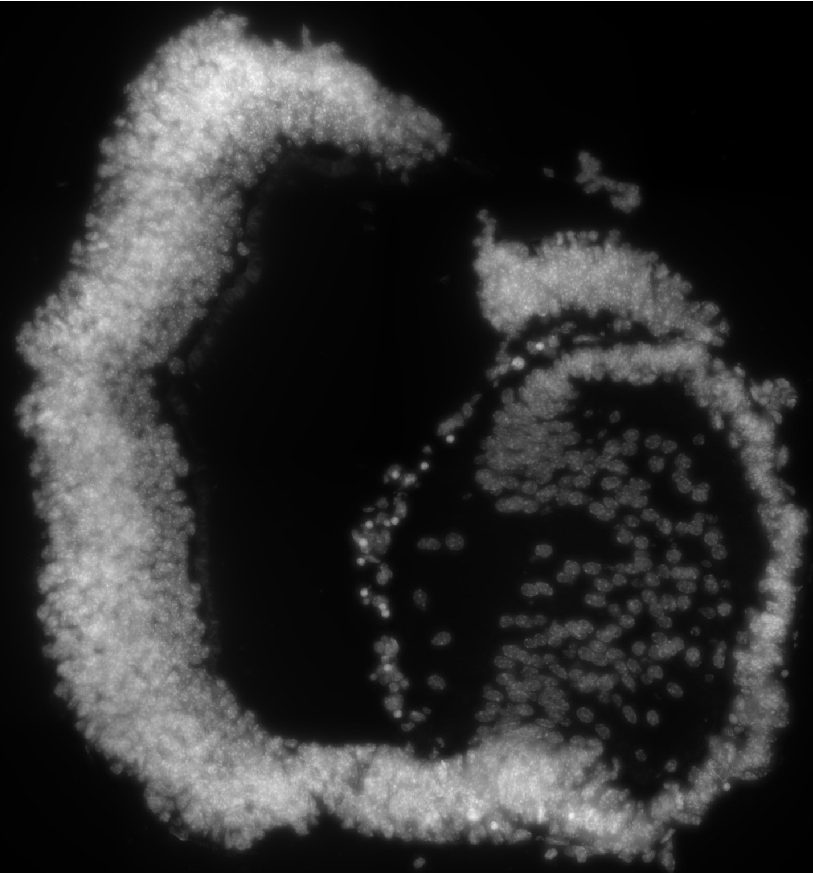}
    \caption{Retina sample}
    \label{fig:third}
\end{subfigure}
        
\caption{Original DAPI images for the three samples}
\label{fig:figures}
\end{figure}

\end{document}